\documentclass{easychair}

\usepackage{latexsym}
\usepackage{amsmath}
\usepackage{amssymb}
\usepackage{stmaryrd}
\usepackage{graphicx}
\usepackage{tikz}
\usepackage{pgfplots}
\usepackage{cite}
\usepackage{booktabs}
\usepackage{adjustbox}
\usepackage{marvosym}
\usepackage{fancyvrb}
\usepackage{bm}

\definecolor{redorange}{rgb}{0.878431, 0.235294, 0.192157}
\definecolor{lightblue}{rgb}{0.95, 0.9, 0.99}
\definecolor{clearyellow}{rgb}{0.964706, 0.745098, 0}
\definecolor{clearorange}{rgb}{0.917647, 0.462745, 0}
\definecolor{mildgray}{rgb}{0.54902, 0.509804, 0.47451}
\definecolor{softblue}{rgb}{0.643137, 0.858824, 0.909804}
\definecolor{bluegray}{rgb}{0.141176, 0.313725, 0.603922}
\definecolor{lightgreen}{rgb}{0.9, 0.99, 0.9}
\definecolor{redpurple}{rgb}{0.835294, 0, 0.196078}
\definecolor{midblue}{rgb}{0, 0.592157, 0.662745}
\definecolor{clearpurple}{rgb}{0.67451, 0.0784314, 0.352941}
\definecolor{browngreen}{rgb}{0.333333, 0.313725, 0.145098}
\definecolor{darkestpurple}{rgb}{0.396078, 0.113725, 0.196078}
\definecolor{greypurple}{rgb}{0.294118, 0.219608, 0.298039}
\definecolor{darkturquoise}{rgb}{0, 0.239216, 0.298039}
\definecolor{darkbrown}{rgb}{0.305882, 0.211765, 0.160784}
\definecolor{midgreen}{rgb}{0.560784, 0.6, 0.243137}
\definecolor{darkred}{rgb}{0.576471, 0.152941, 0.172549}
\definecolor{darkpurple}{rgb}{0.313725, 0.027451, 0.470588}
\definecolor{darkestblue}{rgb}{0, 0.156863, 0.333333}
\definecolor{lightpurple}{rgb}{0.776471, 0.690196, 0.737255}
\definecolor{softgreen}{rgb}{0.733333, 0.772549, 0.572549}
\definecolor{medgreen}{rgb}{0.34, 0.65, 0.34}
\definecolor{offwhite}{rgb}{0.839216, 0.823529, 0.768627}

\newcommand{\leafone}{L_1}
\newcommand{\leafzero}{L_0}

\newcommand{\boolor}{\lor}

\renewcommand{\obar}[1]{\overline{#1}}

\newcommand{\fname}[1]{\mbox{\small\sf #1}}

\newcommand{\var}{\fname{Var}}
\newcommand{\nodes}{\fname{Nodes}}
\newcommand{\bucket}{\fname{BVar}}

\newcommand{\pgbdd}{{\sffamily\scshape pgbdd}}

\newcommand{\pgpbs}{{\sffamily\scshape  pgpbs}}

\newcommand{\kissat}{{\sffamily\scshape kissat}}

\newcommand{\ebddres}{{\sffamily\scshape  ebddres}}
\newcommand{\tbuddy}{{\sffamily\scshape  tbuddy}}

\newcommand{\bddorder}{\pi_{\mathrm{v}}}
\newcommand{\bucketorder}{\pi_{\mathrm{b}}}

\newcommand{\makenode}[1]{\bm{#1}}
\newcommand{\nodeu}{\makenode{u}}
\newcommand{\nodev}{\makenode{v}}
\newcommand{\nodew}{\makenode{w}}

\title{Notes on ``Bounds on BDD-Based Bucket Elimination'' \\ \today}

\author{Randal E. Bryant}
\institute{
  Computer Science Department\\
  Carnegie Mellon University, Pittsburgh, PA, United States\\
  \email{Randy.Bryant@cs.cmu.edu}
}

\authorrunning{R. E. Bryant}

\titlerunning{BDD-Based Bucket Elimination}

\begin{document}

\maketitle

\begin{abstract}
  This paper concerns Boolean satisfiability (SAT) solvers based on Ordered Binary Decision Diagrams (BDDs),
  especially those that can generate proofs of unsatisfiability.
Mengel has presented a theoretical analysis that a BDD-based SAT
solver can generate a proof of unsatisfiability for the pigeonhole
problem (PHP$_n$) in polynomial time, even when the problem is encoded in
the standard ``direct'' form.  His approach is based on {\em bucket
  elimination}, using different orderings for the variables in the
BDDs than in the buckets.  We show experimentally that these proofs
scale as $O(n^5)$.  We also confirm the exponential scaling
that occurs when the same variable ordering is used for the BDDs as for the
buckets.
\end{abstract}

\section{Introduction}

This paper concerns a subclass of Boolean satisfiability (SAT) solvers
that can generate proofs of unsatisfiability when given an
unsatisfiable formula. The subclass uses Ordered Binary Decision
Diagrams (BDDS)~\cite{Bryant:1986} to reason about Boolean formulas.  Examples of proof-generating, BDD-based solvers include
\ebddres~\cite{ebddres,Jussila:2006}, \pgbdd~\cite{bryant:tacas:2021}, and \tbuddy~\cite{bryant:fmcad:2022}.

For the 2023 SAT Conference, Stefan Mengel published insightful
work~\cite{mengel:sat:2023} on the capabilities of {\em bucket
  elimination}~\cite{dechter-ai-1999,pan-sat-2004} as a mechanism for
systematically performing a sequence of conjunction and quantification
operations in a BDD-based SAT solver.  In this note, we experimentally
confirm his analysis that this approach can yield polynomially-sized
unsatisfiability proofs for the {\em pigeonhole problem}, a
well-studied problem for which any resolution proof of its
unsatisfiability must be of exponential length~\cite{Haken:1985},
while there are known proofs of polynomial size using extended
resolution~\cite{Cook:1976}.

Mengel's key insight is that by using different permutations for the
BDD variable ordering and the bucket elimination ordering, the
pigeonhole problem becomes tractable.  We had explored this
possibility in our earlier work on BDD-based SAT
solving~\cite{bryant:tacas:2021} using our proof-generating solver
\pgbdd{}\footnote{Available at {\tt \url{https://github.com/rebryant/pgbdd-artifact}}.}, but we had not tried it on the
pigeonhole problem using the standard ``direct'' encoding.
Instead, we stated in the paper:
\begin{quote}
``Using \pgbdd{},
we were unable to find any strategy that gets beyond $n=16$ with a
direct encoding.  Our best results came from a ``tree'' strategy,
simply forming the conjunction of the input clauses using a balanced tree of
binary operations.''
\end{quote}
Mengel's paper shows that we overlooked a capability built into our solver.
Furthermore, he provides a formal analysis of the complexity.

\section{BDD-Based Bucket Elimination}

Consider a set of BDDs over the variables $X = \{x_1, x_2, \ldots, x_n\}$.
Let $\bddorder$
and $\bucketorder$ denote two permutations of the variables in $X$.
For BDD node $\nodeu$, let $\var(\nodeu)$ denote its associated
variable.
Permutation $\bddorder$ determines the ordering of variables in the
BDD~\cite{Bryant:1986}: if node $\nodeu$ has $\var(\nodeu) = x_i$ and
one of its child nodes $\nodev$ has $\var(\nodev) = x_j$, then these variables must satisfy $\bddorder(x_i) < \bddorder(x_j)$.
For a BDD node
$\nodeu$, let $\nodes(\nodeu)$ denote that set of all nodes occurring
in the subgraph having root $\nodeu$.  By our definitions,
$\var(\nodeu)$ must be the minimum variable, according to $\bddorder$, of all
variables for the nodes in $\nodes(\nodeu)$.

For the BDD with root node $\nodeu$, we let $\bucket(\nodeu)$ denote
the the minimum variable, according to $\bucketorder$, of all
variables for the nodes in $\nodes(\nodeu)$.  That is, when
$\bucket(\nodeu) = x_i$, then 1) there must be some node
$\nodev \in \nodes(\nodeu)$ such that $\var(\nodev) = x_i$, and 2) any node
$\nodew \in \nodes(\nodeu)$ must have its associated variable
$x_j = \var(\nodew)$ satisfy $\bucketorder(x_i) \leq \bucketorder(x_j)$.

Bucket elimination processes a formula in
conjunctive normal form (CNF) as follows.  For each variable $x_i \in X$, we
maintain a set of BDD root nodes $B[x_i]$ (known as a ``bucket''), such
that every node $\nodeu \in B[x_i]$ has $\bucket(\nodeu) = x_i$.
Initially, these buckets are empty.  Each clause is converted into a
BDD by forming the disjunction of its literals, and then its root node $\nodeu$ is placed
in bucket $B[\bucket(\nodeu)]$.

Each bucket is processed in sequence according to the bucket ordering.
Processing bucket $B[x_i]$ involves iterating the following steps until it is empty:
\begin{enumerate}
\item If $B[x_i] = \{\nodeu\}$ for some node $\nodeu$, then remove this node and
  existentially
  quantify $\nodeu$ by variable $x_i$.
  If the resulting node $\nodev$ is not the constant node $\leafone$,
  then place it in bucket $B[\bucket(\nodev)]$.
  We are guaranteed that the destination bucket will be later in the ordering:
  for $\bucket(\nodev) = x_j$, we must have $\bucketorder(x_j) > \bucketorder(x_i)$.
\item If $|B[x_i]| > 1$, first select and remove two nodes $\nodeu$ and
  $\nodev$ from $B[x_i]$ and then form their conjunction $\nodew$.  If
  $\nodew$ is the constant node $\leafzero$, then the formula is
  unsatisfiable.  Otherwise, place $\nodew$ in bucket
  $B[\bucket(\nodew)]$.  In most cases, the destination bucket will be
  $B[x_i]$.  However, it is possible for the conjunction to no longer
  depend on $x_i$.  For example, consider the case where bucket $B[x_1]$ contains BDDs representing
  Boolean formula $x_1 \lor x_2$ and 
  $\obar{x}_1 \lor x_2$.  Their conjunction will be the BDD representation of $x_2$, which should be placed in bucket $B[x_2]$.
  We are guaranteed that the destination bucket will not be earlier in the ordering:
  for $\bucket(\nodew) = x_j$, we must have $\bucketorder(x_j) \geq \bucketorder(x_i)$.
\end{enumerate}
If this process continues without generating constant node
$\leafzero$, then the formula is satisfiable.  Satisfying solutions
can then be generated by assigning values to the variables according
the inverse bucket order~\cite{bryant:fmcad:2022}.

Most implementations of BDD-based bucket
elimination~\cite{bryant:fmcad:2022,Jussila:2006,pan-sat-2004} use the
same ordering for both the BDD variables ($\bddorder$) and bucket
elimination ($\bucketorder$).  This simplifies the task of both
determining the proper bucket for a BDD, since each node $\nodeu$ will have
$\bucket(\nodeu) = \var(\nodeu)$.  It also
simplifies existential quantification: the existential quantification
of $\nodeu$ with respect to $\var(\nodeu)$ is the disjunction of
its two children.  On the other hand, implementing the more generality
capability of distinct orderings is not difficult.  Computing
$\bucket(\nodeu)$ can be done with a simple traversal of the nodes in
$\nodes(\nodeu)$.  The existential quantification of $\nodeu$ by
variable $x_i$ involves computing the restrictions of $\nodeu$ by
$x_i$ and $\obar{x}_i$ and forming their
disjunction~\cite{andersen-1998,bryant-hmc-2018}.  We implemented this
capability in \pgbdd{}~\cite{bryant:tacas:2021}.

\section{Pigeonhole Problem}

The pigeonhole problem is one of the most studied problems in
propositional reasoning.  Given a set of $n$ holes and a set of $n+1$
pigeons, PHP$_{n}$ asks whether there is an assignment of pigeons to holes
such that 1) every pigeon is in some hole, and 2) every hole contains
at most one pigeon.  The answer is no, of course, but any resolution
proof for this must be of exponential length~\cite{Haken:1985}.  
Groote and Zantema have
shown that any BDD-based proof of the principle that only uses the
conjunction operations must be of exponential size~\cite{groote-dam-2003}.
On the other hand, Cook
constructed an extended resolution proof of size $O(n^4)$, in part to
demonstrate the expressive power of extended resolution~\cite{Cook:1976}.

We consider the standard ``direct'' encoding of the problem.
It is based on a set of
variables $p_{i,j}$ for $1 \leq i \leq n$ and $1 \leq j \leq n+1$,
with the interpretation that $p_{i,j}$ true when pigeon $j$ is assigned to hole $i$.
As a running example, we consider the case of $n=2$, having variables $p_{1,1}$ through $p_{2,3}$.

Encoding the property that each pigeon $j$ is assigned to some hole can be
expressed as a single clause:
\begin{eqnarray*}
{\it Pigeon}_j & = & \bigvee_{i = 1}^{n} p_{i,j}
\end{eqnarray*}
For example, with $n=2$, there are three clauses:
\begin{displaymath}
  \begin{array}{lcl}
    {\it Pigeon}_1 &=& p_{1, 1} \lor p_{2, 1}  \\
    {\it Pigeon}_2 &=& p_{1, 2} \lor p_{2, 2} \\
    {\it Pigeon}_3 &=& p_{1, 3} \lor p_{2, 3} \\
  \end{array}
\end{displaymath}

The {\em direct} encoding of
the property that each hole $i$ contains at most one pigeon
simply
states that for any pair of pigeons $j$ and $k$, at least one of them must not
be in hole $i$:
\begin{eqnarray*}
{\it Hole}_i & = & \bigwedge_{j = 1}^{n+1} \bigwedge_{k=j+1}^{n+1} (\obar{p}_{i,j} \boolor \obar{p}_{i,k})
\end{eqnarray*}
This encoding requires $\Theta(n^2)$ clauses for each hole, yielding a total CNF size of $\Theta(n^3)$.
For example, with $n=2$, each hole $i$ requires three clauses:
\begin{displaymath}
  \begin{array}{lcll}
    {\it Hole}_1 &=&   (\obar{p}_{1, 1} \lor \obar{p}_{1, 2}) & \land \\
                  && (\obar{p}_{1, 1} \lor \obar{p}_{1, 3}) & \land \\
                  && (\obar{p}_{1, 2} \lor \obar{p}_{1, 3}) & \\
&&&\\
    {\it Hole}_2  &=& (\obar{p}_{2, 1} \lor \obar{p}_{2, 2}) & \land \\
                  & & (\obar{p}_{2, 1} \lor \obar{p}_{2, 2}) & \land \\
                  & & (\obar{p}_{2, 1} \lor \obar{p}_{2, 2}) & \\
  \end{array}
\end{displaymath}

We consider two different orderings of the encoding variables.
The {\em pigeon-major} ordering lists all the
variables for each pigeon in succession.  That is, it first lists the
variables $p_{1,1}, p_{1,2}, \ldots, p_{1, n+1}$, and then the
variables $p_{2,1}, p_{2,2}, \ldots, p_{2, n+1}$, and so on, finishing
with variables $p_{n,1}, p_{n,2}, \ldots, p_{n, n+1}$.  
For example, with $n=2$, the pigeon-major order is:
\begin{displaymath}
p_{1,1},\; p_{1,2},\; p_{1,3},\; p_{2,1},\; p_{2,2},\; p_{2, 3}
\end{displaymath}

The {\em
  hole-major} ordering lists the variables for each hole in
succession.  That is, it first lists the variables $p_{1,1}, p_{2,1},
\ldots, p_{n,1}$, and then the variables $p_{1,2}, p_{2,2}, \ldots,
p_{n,2}$, and so on, finishing with variables $p_{1,n+1}, p_{2,n+1},
\ldots, p_{n,n+1}$.
For example, with $n=2$, the hole-major order is:
\begin{displaymath}
p_{1,1},\; p_{2,1},\; p_{1,2},\; p_{2,2},\; p_{1,3},\; p_{2,3}
\end{displaymath}

We refer to these two orderings as being {\em orthogonal}: by viewing
the variables $p_{i,j}$ as entries in a rectangular matrix, pigeon-major ordering
corresponds to a row-major ordering, while hole-major corresponds to
column-major.

\begin{figure}
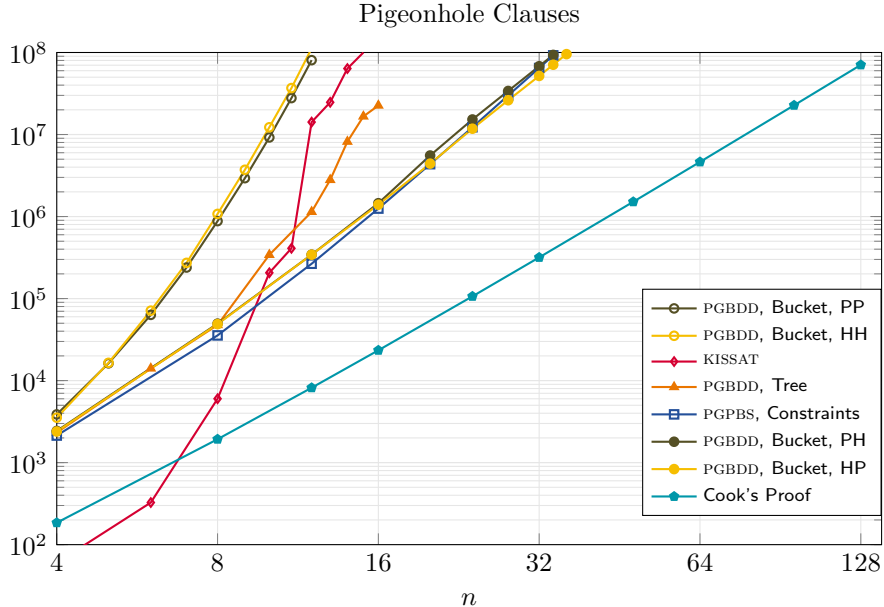

\centering
\begin{tikzpicture}[scale = 1.0]
          \begin{axis}[mark options={scale=0.8},grid=both, grid style={black!10}, xmode=log, ymode=log, legend style={at={(1.00,0.52)}}, legend cell align={left},
                              x post scale=1.6, y post scale=1.15, xlabel=$n$, xtick={4,8,16,32,64,128, 256}, xticklabels={$4$,$8$,$16$,$32$,$64$,$128$},xmin=4,xmax=140,ymin=100,ymax=100000000, title={Pigeonhole Clauses}]
            \input{pigeon-bucket-pp}
            \input{pigeon-bucket-hh}
            \input{pigeon-direct-cdcl}
            \input{pigeon-direct-pm-linear}
            \input{pigeon-average-constraint-randomorder}
            \input{pigeon-bucket-ph}
            \input{pigeon-bucket-hp}
            \input{pigeon-direct-cook}
 
            \legend{
              \scriptsize \textsf{\pgbdd, Bucket, PP},
              \scriptsize \textsf{\pgbdd, Bucket, HH},
              \scriptsize \textsf{\kissat},
              \scriptsize \textsf{\pgbdd, Tree},
              \scriptsize \textsf{\pgpbs, Constraints},
              \scriptsize \textsf{\pgbdd, Bucket, PH},
              \scriptsize \textsf{\pgbdd, Bucket, HP},
              \scriptsize \textsf{Cook's Proof},
            }
          \end{axis}
\end{tikzpicture}
\caption{Total number of clauses in proofs of pigeonhole problem for $n$ holes.  All are based on a direct encoding}
\label{fig:data:pigeonhole}
\end{figure}  

\section{Experimental Results}

Figure~\ref{fig:data:pigeonhole} shows our measurements for the sizes
of the unsatisfiability proofs (measured as the number of input and
proof clauses), as a function of $n$, for different proof generation
methods.  In each case, we show how large $n$ can be before the proof
exceeds $10^8$ clauses.  The plot labeled \kissat{} is for the CDCL
solver \kissat{}~\cite{biere-kissat-2020}, considered to be the state-of-the art for SAT
solvers.  Since the steps taken by CDCL solvers can be encoded as
resolution proofs, we expect the proof sizes to grow exponentially,
and this is borne out here.  It cannot go beyond $n = 14$ within the
clause limit.

Similarly, our previous attempt with \pgbdd{}, labeled ``Tree''
showed exponential growth.  For this, we simply formed the conjunction
of the BDDs for the clauses using a binary tree of conjunction
operations.  In 2022 we presented \pgpbs{}, a BDD-based solver for
pseudo-Boolean constraints~\cite{bryant:tacas:2022}.  This solver
can convert a clausal representation of the problem into one
consisting of integer linear constraints, where each variable must be
assigned value 0 or 1.  It then uses Fourier-Motzkin
elimination~\cite{dantzig:1974,williams:1975} to combine the
constraints and show that they are infeasible.  This program
achieves polynomial performance on the problem, scaling as $O(n^5)$.

The four plots labeled ``\pgbdd{}, Bucket'' show the result for
running \pgbdd{} with bucket elimination using four different combinations
of permutations of the variables $p_{i,j}$ for the variable and bucket
orderings.  
The notation in the legend  first lists the
ordering used for bucket elimination (``P'' for pigeon-major and ``H'' for hole-major) 
and then the ordering for the
BDD variables.  

As can be seen, the two cases where the same ordering is used for both
the buckets and the BDDs (PP and HH) have exponential growth.  They
cannot even match the performance of \kissat{}.  This confirms the
lower bound proved by Mengel~\cite{mengel:sat:2023} for what he refers
to as ``single-order bucket elimination.''  On the other hand, when
the bucket ordering is orthogonal to that of the BDDs, then the
performance closely matches the $O(n^5)$ scaling seen for \pgpbs{}.
Although it is hard to distinguish the three plots, the best overall
performance comes from bucket elimination using a hole-major bucket
ordering and a pigeon-major variable ordering.  For this version, the
proof for $n=36$ contains 95,494,509 clauses, the only one that is
below $10^8$ clauses for this value of $n$.

\section{Conclusion}

The problem of selecting a proper variable ordering for BDDs has been
a source of study and frustration for decades.  To this, we now add
the task of selecting the proper bucket ordering.  As Mengel's
analysis shows, the interactions between these two orderings can be
quite subtle and lead to surprising results.

Cook's proof scales as $O(n^4)$, asymptotically better than any we
have generated with BDDs.  Grosof, Zhang, and Heule have constructed a proof
with $O(n^3)$
clauses~\cite{grosof:pos:2022}.  This is optimal, since the problem
representation itself requires $\Theta(n^3)$ clauses.  It would be
interesting to find an automated algorithm that could improve on our
$O(n^5)$ scaling.

\bibliography{references}

\begin{thebibliography}{10}

\bibitem{andersen-1998}
H.~R. Andersen.
\newblock An introduction to binary decision diagrams.
\newblock Technical report, Technical University of Denmark, October 1997.

\bibitem{biere-kissat-2020}
A.~Biere, K.~Fazekas, M.~Fleury, and M.~Heisinger.
\newblock {CaDiCaL}, {Kissat}, {Paracooba}, {Plingeling} and {Treengeling}
  entering the {SAT Competition 2020}.
\newblock In {\em Proc.~of {SAT Competition} 2020---Solver and Benchmark
  Descriptions}, volume B-2020-1 of {\em Department of Computer Science Report
  Series B}, pages 51--53. University of Helsinki, 2020.

\bibitem{Bryant:1986}
R.~E. Bryant.
\newblock Graph-based algorithms for {B}oolean function manipulation.
\newblock {\em IEEE Trans. Computers}, 35(8):677--691, 1986.

\bibitem{bryant-hmc-2018}
R.~E. Bryant.
\newblock Binary decision diagrams.
\newblock In E.~M. Clarke, T.~A. Henzinger, H.~Veith, and R.~Bloem, editors,
  {\em Handbook of Model Checking}, pages 191--217. Springer, 2018.

\bibitem{bryant:fmcad:2022}
R.~E. Bryant.
\newblock {TBUDDY}: A proof-generating {BDD} package.
\newblock In {\em Formal Methods in Computer-Aided Design (FMCAD)}, 2022.

\bibitem{bryant:tacas:2022}
R.~E. Bryant, A.~Biere, and M.~J.~H. Heule.
\newblock Clausal proofs for pseudo-{B}oolean reasoning.
\newblock In {\em Tools and Algorithms for the Construction and Analysis of
  Systems (TACAS)}, volume 12651 of {\em LNCS}, pages 76--93, 2022.

\bibitem{bryant:tacas:2021}
R.~E. Bryant and M.~J.~H. Heule.
\newblock Generating extended resolution proofs with a {BDD}-based {SAT}
  solver.
\newblock In {\em Tools and Algorithms for the Construction and Analysis of
  Systems (TACAS), Part I}, volume 12651 of {\em LNCS}, pages 76--93, 2021.

\bibitem{Cook:1976}
S.~A. Cook.
\newblock A short proof of the pigeon hole principle using extended resolution.
\newblock {\em SIGACT News}, 8(4):28--32, Oct. 1976.

\bibitem{dantzig:1974}
G.~B. Dantzig and B.~C. Eaves.
\newblock {F}ourier-{M}otzkin elimination and its dual with application to
  integer programming.
\newblock In {\em Combinatorial Programming: Methods and Applications}, pages
  93--102. Springer, 1974.

\bibitem{dechter-ai-1999}
R.~Dechter.
\newblock Bucket elimination: A unifying framework for reasoning.
\newblock {\em Artificial Intelligence}, 113(1--2):41--85, 1999.

\bibitem{groote-dam-2003}
J.~F. Groote and H.~Zantema.
\newblock Resolution and binary decision diagrams cannot simulate each other
  polynomially.
\newblock {\em Discrete Applied Mathematics}, 130(2):157--171, 2003.

\bibitem{grosof:pos:2022}
I.~Grosof, N.~Zhang, and M.~J.~H. Heule.
\newblock Toward the shortest {DRAT} proof of the pigeonhole principle.
\newblock In {\em Pragmatics of SAT}, 2022.

\bibitem{Haken:1985}
A.~Haken.
\newblock The intractability of resolution.
\newblock {\em Theoretical Computer Science}, 39:297--308, 1985.

\bibitem{Jussila:2006}
T.~Jussila, C.~Sinz, and A.~Biere.
\newblock Extended resolution proofs for symbolic {SAT} solving with
  quantification.
\newblock In {\em Theory and Applications of Satisfiability Testing (SAT)},
  volume 4121 of {\em LNCS}, pages 54--60, 2006.

\bibitem{mengel:sat:2023}
S.~Mengel.
\newblock Bounds on {BDD}-based bucket elimination.
\newblock In {\em Theory and Applications of Satisfiability Testing (SAT)},
  2023.

\bibitem{pan-sat-2004}
G.~Pan and M.~Y. Vardi.
\newblock Search vs.~symbolic techniques in satisfiability solving.
\newblock In {\em Theory and Applications of Satisfiability Testing (SAT)},
  volume 3542 of {\em LNCS}, pages 235--250, 2005.

\bibitem{ebddres}
C.~Sinz and A.~Biere.
\newblock Extended resolution proofs for conjoining {BDD}s.
\newblock In {\em Computer Science Symposium in Russia (CSR)}, volume 3967 of
  {\em LNCS}, pages 600--611, 2006.

\bibitem{williams:1975}
H.~P. Williams.
\newblock {F}ourier-{M}otzkin elimination extension to integer programming
  problems.
\newblock {\em Journal of Combinatorial Theory (A)}, 21:118--123, 1976.

\end{thebibliography}

\end{document}